# Calculation of the Electronic Structure and Optical Absorption of the Ligated Magic-Size Clusters with $Cd_8Se_{13}$ Core

V. S. Gurin

*Research Institute for Physical Chemical Problems, Belarusian State University,*

*Leningradskaya 14, 220006, Minsk, Belarus*

**Abstract**

A series of the cadmium selenide magic size clusters with the cores of $Cd_8Se_{13}$ terminated with H- and N-containing ligands are calculated at the DFT level with different functionals and ECP basis sets. The optimum functional was selected on the basis of calculation for diatomic CdSe and other smallest clusters with available experimental data and the higher-level calculations. Optical absorption spectra are simulated through the TDDFT method. The results on the ligand effects are in consistence with experimental data for the magic size clusters produced in pyridine medium.

## 1. Introduction

Cadmium chalcogenide nanoclusters are of great interest last years in line with nanosized materials of other semiconductor compounds. CdSe and CdS may be considered as reference compounds for both experimental and theoretical studies in this field and intensive studies of CdSe and CdS nanoclusters is continued to date [1-3]. A special interest can be noted for the nanoclusters with predefined number of atoms, so called magic size clusters (MSC). They can provide the knowledge of exact atomic structure rather than just geometric size and results in the detailed information on their electronic structure and properties. A number of MSC of various compositions has been successfully prepared through the elaborated protocols, characterized using different spectroscopic techniques and treated theoretically interpreting the properties through electronic structure calculations. The examples of these clusters are rather numerous, they include the cores with $Cd_4$, $Cd_8$, $Cd_{10}$, $Cd_{13}$, $Cd_{17}$, $Cd_{28}$, $Cd_{33}$, etc. [2,3]. Among these MSC of various composition and structure, clusters of the lower size range are of interest as identification of the early steps of nucleation of nanoparticles and quantum dots those determine the pathways of their growth, 1D, 2D, 3D, and various nanocrystalline phases. In the present work, we concern the series of $Cd_8$- species those were treated considerably rare to date. An experimental verification of the MSC of this size is also not very easy because their extreme reactivity, rapid growth and strong deviation from the stoichiometric Cd/Se=1/1 ratio. Meanwhile, there is experimental information on the MSC with $Cd_8Se_{13}$- core stabilized by N-terminated pyridine derivatives [4]. Calculation of the electronic structure of these MSC was not be done to date for our best knowledge. However, a theoretical treatment of them is of great importance for understanding their properties and relation with another series of CdSe MSC. This $Cd_8$-series is featured by the geometry built with Se atom in the centre (Fig. 1). In the present study we performed the DFT level caculations of the geometry and electronic structure of the MSC with $Cd_8Se_{13}$-core terminated by hydrogen, amine ligands and pyridine group. Recently, a part of theoretical results has been published for H- and amine- terminated $Cd_8Se_{13}$ and $Hg_8Se_{13}$ MSC [5]. Before the calculation of these MSC, we pay more attention on justification of functional and basis set with the best correspondence to experimental and high-level theoretical data for the simpler systems, CdSe diatomic, $Cd_2Se_2$ molecules, and the ligated $Cd_2Se_2$.

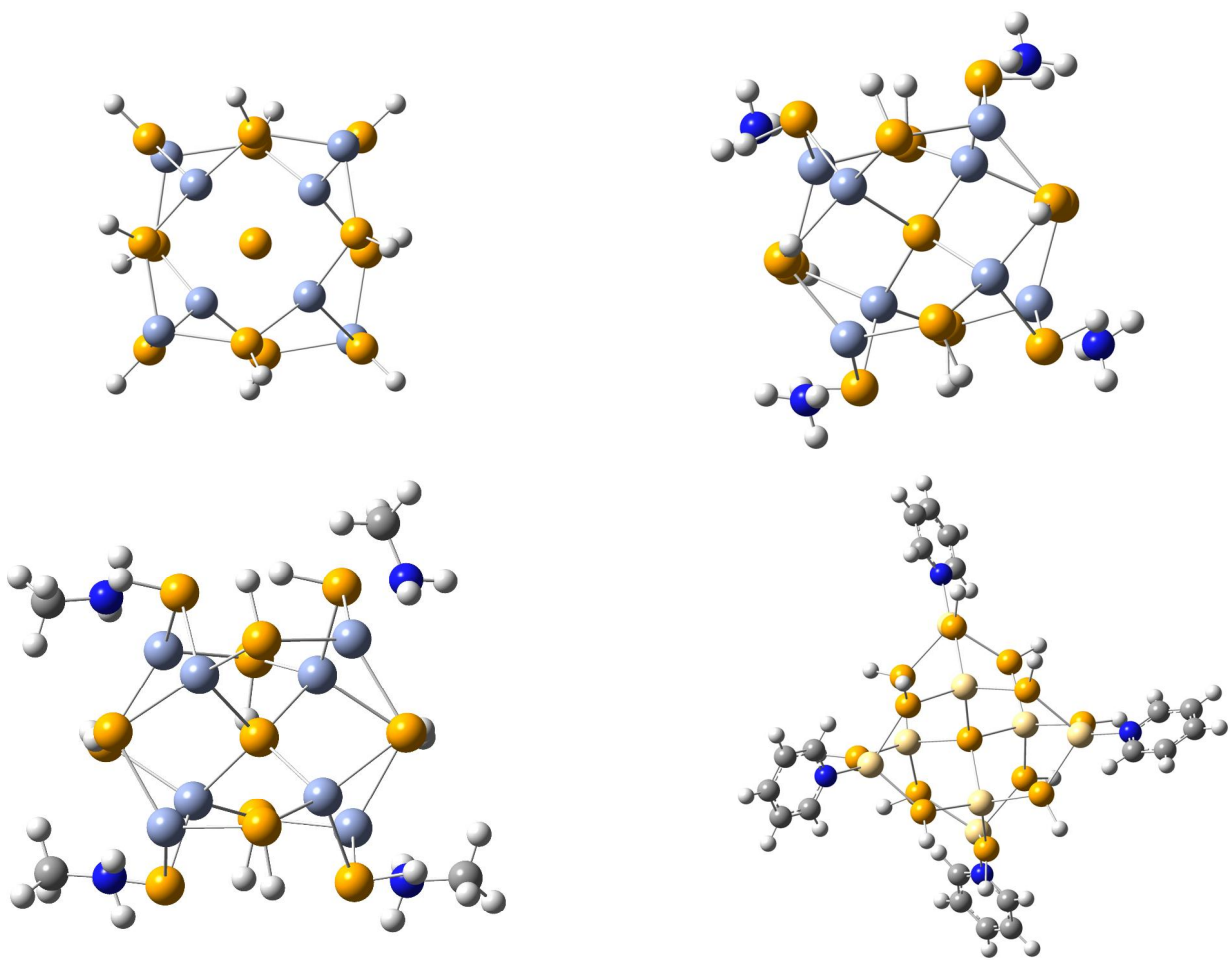

Fig. 1 Cluster models based on $Cd_8Se_{13}$-core with different ligands used for calculations: –H, -NH$_3$, -NH$_2$-CH$_3$, and pyridine (NC$_5$H$_5$).

## 2. Building of the Cd$_8$Se$_{13}$ structure and calculation technique

The cluster structures under calculation within the framework of the present study were built starting from the central selenium atom (Fig. 1) assuming its tetrahedral environment, Se--Cd$_4$. Further, to each Cd atoms next Se atoms (3Se to one Cd) were attached resulting in Se—[Cd-Se$_3$]$_4$ = Cd$_4$Se$_{13}$, and finally, else four Cd were attached closing the dangling Se atoms. Hydrogens were linked to the surface Se atoms. Thus, the structure of composition Cd$_8$Se$_{13}$H$_{12}$ is obtained with a full charge +2.

The ligand-bound models were terminated with nitrogen-containing groups, NH$_3$, and NH$_2$CH$_3$ those can be considered as simplest simulation of pyridine molecules, providing two principal aspects: M-N bonds and -M-N-C- sequence at the cluster surface. In the fourth model, pyridine molecules were linked by N atoms like the above amines (Fig. 1).

Calculations were done using DFT (density functional theory) with extension to time-dependent version (TDDFT) that allows get information both on geometry and electronic transitions of the cluster models with many atoms of heavy elements at reasonable computational resources. In spite of many calculations published to date at this and higher level of theory for various composition cadmium chalcogenide clusters [6-29], there is noticeable variance in details of calculations (type of functional, effective core potentials (ECP), basis sets, etc.) resulting in slightly different data both for geometrical parameters and electronic structure those cannot be directly verified in experiment. Optical absorption spectra simulated through electronic transitions between molecular orbitals in clusters can be compared with available experimental data, but there is also variance within the expectable range of transition energies derived by different calculations [6-13,16-24,26-33]. Thus, an optimum choice of calculation details required special care to get most adequate results.

In order to adjust the calculation technique within the framework of DFT level several standard functionals and basis sets with ECP have been tested comparing the calculation data with corresponding results for the simpler species than the models for MSC Cd$_8$-selenide clusters. These are CdSe diatomic molecules, Cd$_2$Se$_2$ molecules, and complex compounds with Cd$_2$Se$_2$ cores that include metal-nitrogen bonds and hydrocarbon moiety, Cd$_2$Se$_2$-(NH$_2$-CH$_2$-CH$_2$-CH$_3$). Crystallographic data available for these molecules (Table 1) are used for comparison with calculations. It collects the selected data of these calculations which justify the accuracy of usage the combined exchange-correlation functional denoted as XA-SVWN5 (in GAUSSIAN terms) while the popular functional of the B3LYP type resulted in the essentially longer values for bond lengths. The ECP basis set of the CEP-121g type (also in GAUSSIAN terms) provides the calculations taking into account the valence electronic shells of Cd (34 basis functions) and Se (8 basis functions) retaining 28e cores for Cd and Se. This choice of functional and basis set results in good agreement with the data for distances in the diatomics and $Cd_2Se_2$- test compounds. Reference values for binding energies (both experimental and theoretical) appear to be not too close to the calculated ones since a great deviation exists in these data because ambiguous assignment of the ground electronic states of the molecules and lack of experimental information. In the case of complex compounds with $Cd_2Se_2$ cores we used the comparison with crystallographic data for solid phases, composition of which is rather close to the calculated molecules (Table 1). The data for $Cd_2Se_2$-aminopropyl indicate that our XA-SVWN5/CEP-121g calculations provide better correspondence of distances with the values for crystals, in particular, Cd-Se and Cd-N (boldfaced in Table 1; the shorter second distance 2.513 Å appears due to symmetry breaking since aminopropyl group is bound at one side of molecule while in crystals the symmetry is closer to rhombus by a neighbor sublattice).

Electronic transitions between molecular orbitals of the clusters $M_8Se_{13}H_{12}$-R (R=$NH_3$, $NH_2$-$CH_3$, Pyr) with optimized geometries were calculated for simulation of absorption spectra using TDDFT technique covering excited states of the clusters in the energy range up to 5-6 eV. For TDDFT calculations ECP basis sets of LANL2DZ type and B3LYP functional were used. Such choice of details in TDDFT demonstrates adequate results for CdSe clusters [12,13,16-19,26-35] together with reasonable computing resources to calculate 50-100 excited states. All calculations of this work were done with GAUSSIAN09 package [36].

## 3. Calculation results

Numerical data for geometrical parameters of the clusters under consideration are given in Table 1. The clusters are depicted in Fig. 1 using final optimized atomic coordinates. The data for geometry of clusters indicate that the distances Cd-Se enter the reasonable range, slightly shorter than corresponding bulk counterparts for sphalerite CdSe lattice (2.631 Ang, the unit cell constant a=6.077 Ang) and essentially longer than in diatomic CdSe molecules (referenced to the calculations of more high-level methods and experiment). Because of the effect of dangling atoms and groups -H, -$NH_3$, -$NH_2CH_3$, and pyridine (-$NC_5H_5$) upon structure of the clusters, the point group symmetry of initial cores ($T_d$ in the case of perfect building of models) is lost, however, the metal selenide cores retain fragments of the sphalerite lattice. Due to little difference in the distances of internal and external parts this lattice fragments appear as distorted sphalerite. A general trends in changes of Cd-Se distances are the following: (i) interior bonds are shorter than external ones and this difference is more explicit in the case of $Cd_8$- than $Cd_{17}$- (presented in Ref.5); (ii) the termination with hydrogen results in essentially non-uniform distribution in the Cd-Se bond lengths; (iii) Cd-N distances appear to be the longer for smaller ligands, and enter the range typical for Cd-N distances in chelate compounds, e.g. $Cd^{2+}$ with cysteine [37]. This strong interaction can be associated with principal role of N-active ligands in the stabilization of this type of MSC [4,5].

The energy gap between the frontier orbitals, HOMO and LUMO, characterizes features of the clusters, and in the case of series with increasing number of atoms (cluster sizes) this value usually reveals regular behavior or odd-even oscillations in dependence on the number of atoms when a cluster series is long. In our case, the clusters under study are different by the ligand types rather than the number of atoms. The HOMO-LUMO gap for the clusters with N-containing ligands can be considered as approaching to energies of electronic transitions (below, TDDFT results), i.e. the transitions can be associated with the near-gap orbitals. The effect of the termination type upon the energy gap is rather large that may be understood because small size of this cluster series with very essential contribution of electronic density redistribution passing from H- to the other ligands.

The calculated absorption spectra are presented in Fig. 2 as positions of electronic transitions in the wavelength scale indicating relative intensities derived from the corresponding oscillator forces. This series of the spectra falls into the visible and UV ranges similar to various experimental ones for CdSe MSC of different composition [19-24] and theoretical spectra for simulated MSC $M_nX_m$ (n,m < 100) [7-33]. Meanwhile, all absorption maxima are at the shorter wavelength range than the ranges for typical ligand-stabilized CdSe quantum dots with sizes 1-2 nm and more those are not usually assigned to MSC. The effect of different ligands upon the electronic transitions calculated is rather significant. That is quite expectable for the clusters under study those are small and they scarcely may be treated in the terms of internal and external shells.

The first absorption line simulated for H-terminated cluster, $Cd_8Se_{13}H_{12}^{2+}$, is at 404 nm and ligation with ammonia results in the noticeable blue shift for the first absorption line. The heavier ligand, $NH_2CH_3$- reverts the blue shift, and for pyridine it is again about 400 nm. The experimental value for the first absorption maximum in the situation closest to this model, the MSC with identified $Cd_8Se_{13}$ core in pyridine medium, is 320 nm. This can be considered as sufficiently good agreement since some secondary effects were not treated in the model: solvent effect, charge of the cluster, possible association, etc. There is feasibility to take into account more effects in the future improved models.

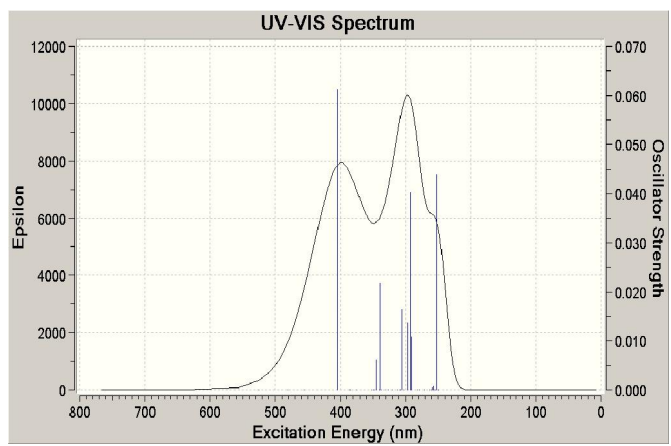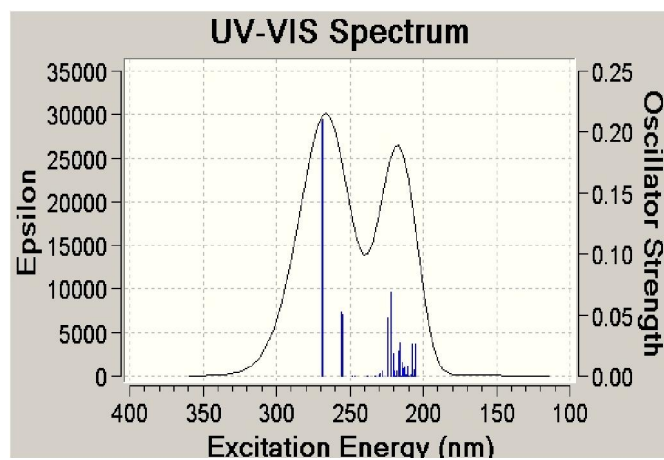
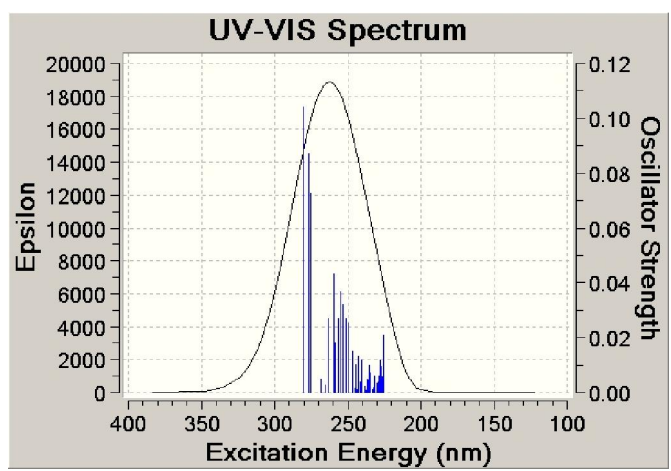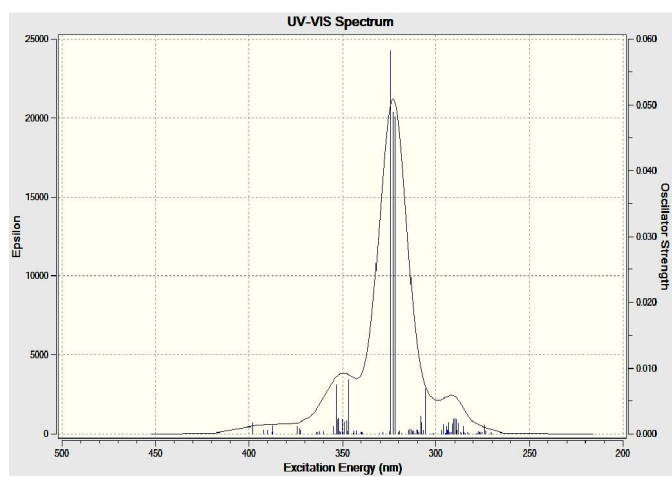

Fig. 2. Absorption spectra calculated for the cluster with different ligands:
–H, -$NH_3$, -$NH_2$-$CH_3$, and pyridine -$NC_5H_5$.

Table 1. Data for CdSe molecules and other smallest particles

| CdSe molecule | | | | | | | |
|---|---|---|---|---|---|---|---|
| Electronic state | Method | ECP | Basis | Functional | $-E_{total}$, $R_{Cd-Se}=R_e$, a.u. | $E_b$, eV [*] | $R_e$, Å |
| $^1\Sigma^+$ | DFT | LANL2DZ | LANL2DZ | B3LYP | 57.2705 | 2.024 | 2.4751 |
| $^3\Pi$ | DFT | LANL2DZ | LANL2DZ | B3LYP | 57.2747 | 0.563 | 2.7762 |
| $^1\Sigma^+$ | DFT | CEP-121g | CEP-121g | B3LYP | 176.2559 | 2.169 | 2.4303 |
| $^3\Pi$ | DFT | CEP-121g | CEP-121g | B3LYP | 176.2586 | 0.520 | 2.6953 |
| **$^1\Sigma^+$** | **DFT** | **CEP-121g** | **CEP-121g** | **XA-SVWN5** | **176.6755** | **3.050** | **2.3521** |
| **$^3\Pi$** | **DFT** | **CEP-121g** | **CEP-121g** | **XA-SVWN5** | **176.6659** | **1.140** | **2.5803** |
| $^1\Sigma^+$ | DFT | LANL2DZ | LANL2DZ | XA-SVWN5 | 57.7097 | 2.928 | 2.4101 |
| $^3\Pi$ | DFT | LANL2DZ | LANL2DZ | XA-SVWN5 | 57.7017 | 1.072 | 2.6794 |
| $^1\Sigma^+$ | DFT | SDD | SDD | B3LYP | 177.1218 | 2.055 | 2.4201 |
| $^3\Pi$ | DFT | SDD | SDD | B3LYP | 177.1246 | 0.5496 | 2.6946 |
| $^1\Sigma^+$ | DFT | SDD | SDD | XA-SVWN5 | 177.5175 | 3.072 | 2.3429 |
| $^3\Pi$ | DFT | SDD | SDD | XA-SVWN5 | 177.5080 | 1.113 | 2.5769 |
| $^1\Sigma^+$ | DFT | LANL2DZ | LANL2DZ | CAM-B3LYP | 57.1681 | 1.916 | 2.4514 |
| $^3\Pi$ | DFT | LANL2DZ | LANL2DZ | CAM-B3LYP | 57.1773 | 0.566 | 2.7131 |
| $^1\Sigma^+$ | CCSD | SDD | SDD | | 176.1122 | 1.891 | 2.4135 |
| $^3\Pi$ | CCSD | SDD | SDD | | 176.1379 | 0.811 | 2.6455 |
| $^1\Sigma^+$ | CCSD | LANL2DZ | LANL2DZ | | 55.6671 | 1.627 | 2.4335 |
| $^3\Pi$ | CCSD | LANL2DZ | LANL2DZ | | 55.6946 | 0.612 | 2.6743 |
| $^1\Sigma^+$ | CCSD | CEP-121g | CEP-121g | | 175.3695 | 1.777 | 2.4252 |
| $^3\Pi$ | CCSD | CEP-121g | CEP-121g | | 175.3970 | 0.737 | 2.6474 |
| $^1\Sigma^+$ | CISD | LANL2DZ | LANL2DZ | | 55.6616 | 1.540 | 2.4323 |
| $^1\Sigma^+$ | QCISD | LANL2DZ | LANL2DZ | | 55.6687 | 1.649 | 2.4308 |

[*]$E_b$ was evaluated for the dissociation to: $Cd^0(^1S) + Se^0(^1D)$ for $^1\Sigma^+$ state and $Cd^0(^1S) + Se^0(^3P)$ for $^3\Pi$, respectively.

| References | Electronic state | Method/Basis set | $E_b$, eV [*] | $R_e$, Å |
|---|---|---|---|---|
| **Experiment** | | | | |
| Huber K P, Herzberg G. Molecular Spectra and Molecular structure. IV. Constants of Diatomic Molecules, 1979, | | | ≤1.99 | |
| CRC Handbook on Chemistry and Physics 2015-2016 (Ed. D.R.Lide), CRC Press LLC referred to: Grade M., Hirschwald W. Bericht. Bunsenges. Phys. Chem. 86 (1982) 899 | | | 1.32 | |
| **Theory** | | | | |
| Peterson K A et al Mol. Phys. 105 (2007) 1139-55 | $^1\Sigma^+$ | CCSD(T)/AVTZ | 1.82 | **2.3726-** |
| | $^1\Sigma^+$ | MRCI+Q/AVTZ | 1.82 | **2.3724** |
| | $^1\Sigma^+$ | MRCI+Q/CBS | 2.00 | **2.361** |
| | $^3\Pi$ | MRCI+Q/CBS | 0.68 | **2.555** |
| Karamanis P. et al J. Chem. Phys. 124 (2006) 071101 | $^1\Sigma^+$ | MP2/3-21G | no data | 2.4003 |

| Reference | Electronic state Symmetry | Method/Basis | $E_b$ | $R_{Cd-Se}$, Å |
|---|---|---|---|---|
| Sen S. et al Phys. Rev. B74 (2006) 205435 | $^1\Sigma^+$ | B3LYP/3-21G | no data | 2.3926 |
| Deglmann P. et al J. Chem. Phys. 116 (2002) 1585 | $^1\Sigma^+$ | BP86/TZVPP | 0.875 | 2.379 |
|  | $^3\Pi$ | BP86/TZVPP | 0.918 | 2.425 |

**$Cd_2Se_2$**

| Electronic state Symmetry | Method | ECP | Basis | Functional | $E_{total}$ ground state, a.u. | $E_b^*$, eV | $E_{HOMO-LUMO}$, eV | $R_{Cd-Se}$, Å | $R_{Cd-Cd}$, Å | Bond angles Se-Cd-Se, deg |
|---|---|---|---|---|---|---|---|---|---|---|
| $^1A_g$ $D_{2h}$ | **DFT** | **CEP-121g** | **CEP-121g** | **XA-SVWN5** | **353.5028** | **6.933** **4.131** | **1.224** | **2.545** | **2.728** | **115.18** |
| $^1A_g$ $D_{2h}$ | DFT | LANL2DZ | LANL2DZ | B3LYP | 114.6576 | 7.222 3.173 | 2.177 | 2.616 | 2.981 | 111.88 |
| $^1A_g$ $D_{2h}$ | DFT | CEP-121g | CEP-121g | B3LYP | 352.6255 | 7.236 4.142 | 2.231 | 2.617 | 2.874 | 113.39 |
| $^1A_g$ $D_{2h}$ | CCSD | LANL2DZ | LANL2DZ |  | 111.5032 | 7.848 4.599 | 7.325 | 2.606 | 2.842 | 113.90 |
| $^1A_g$ $D_{2h}$ | CCSD | CEP-121g | CEP-121g |  | 350.9040 | 8.044 3.554 | 7.310 | 2.604 | 2.843 | 113.82 |
| $^1A_g$ $D_{2h}$ | CCSD | SDD | SDD |  | 352.3571 | 7.394 3.782 | 9.965 | 2.593 | 2.832 | 113.80 |

*$E_b$ was evaluated for the dissociation to: $2[Cd(^1S) + Se(^1D)]$ - 1st line, and $2CdSe(^1\Sigma^+)$ - 2nd line, respectively.

| References | Electronic state Symmetry | Method/Basis set | $E_b$, eV | $R_{Cd-Se}$, Å | $R_{Cd-Cd}$, Å | Bond angles Se-Cd-Se, deg |
|---|---|---|---|---|---|---|
| **Theory** |  |  |  |  |  |  |
| Sen S. et al Phys.Rev.B74 (2006) 205435 |  | B3LYP/3-21G |  | 2.5918 |  | 111.96 |
| Deglmann P. et al J.Chem.Phys. 116 (2002) 1585 |  | MP2/TZVPP | 4.91 | 2.518 |  | 113.3 |
|  |  | BP86/TZVP | 4.95 | 2.591 |  | 113.5 |
|  |  | BP86/TZVPP | 5.06 |  |  |  |
|  |  | B3LYP/TZVP | 4.30 | 2.592 |  | 112.9 |
| Yang P. et a J. Chem. Phys. **129** (2008) 074709 * |  | B3LYP/LANL2DZ |  | 2.641 | 2.947 |  |
|  |  | B3LYP/SDD |  | 2.582 | 2.815 |  |
|  |  | OPBE/LANL2DZ |  | 2.631 | 2.891 |  |
|  |  | OPBE/SDD |  | 2.563 | 2.746 |  |
|  |  | SVWN5/SDD |  | 2.538 | 2.735 |  |
| Karamanis P. et al J.Chem.Phys. 124 (2006) 071101 |  |  |  |  |  |  |
| Jager M. et al J. Chem. Phys. 149 (2018) 244308 |  | PBE0/cc-pVTZ-PP |  | 2.55 |  |  |

| Cd$_2$Se$_2^+$ Rhombus | | | | | | | | | | |
|---|---|---|---|---|---|---|---|---|---|---|
| Electronic state Symmetry | Method | ECP | Basis | Functional | E$_{total}$ ground state, a.u. | E$_b$*, eV | E$_{HOMO-LUMO}$, eV | R$_{Cd-Se}$, Å | R$_{Cd-Cd}$, Å | Bond angles Se-Cd-Se, deg |
| $^2B_{2g}$ D$_{2h}$ | **DFT** | **CEP-121g** | **CEP-121g** | **XA-SVWN5** | **353.1988** | **11.494** | **1.622(α) 0.578(β)** | **2.556** | **2.842** | **112.45** |
| $^2B_{2g}$ D$_{2h}$ | DFT | LANL2DZ | LANL2DZ | B3LYP | 114.3753 | 8.354 | 2.106(α) 1.209(β) | 2.680 | 3.173 | 107.41 |
| $^2B_{2g}$ D$_{2h}$ | DFT | CEP-121g | CEP-121g | B3LYP | 352.3368 | 5.230 | 3.249(α) 1.267(β) | 2.628 | 3.030 | 109.59 |

*E$_b$ was evaluated for the dissociation to: Cd($^1$S) + Cd$^+$($^2$S) +2Se($^1$D)

| References | Electronic state Symmetry | Method/Basis set | E$_b$, eV | R$_{Cd-Se}$, Å | R$_{Cd-Cd}$, Å | R$_{Se-Se}$, Å | Bond angles Se-Cd-Se, deg |
|---|---|---|---|---|---|---|---|
| **Theory** | | | | | | | |
| Jager M. et al J. Chem. Phys. 149 (2018) 244308 | | PBE0/cc-pVTZ-PP | | 2.59 2.53** | | | |

**rhombus is distorted

| Cd$_2$Se$_2^+$ Trapecium | | | | | | | | | | |
|---|---|---|---|---|---|---|---|---|---|---|
| Electronic state Symmetry | Method Symmetry | ECP | Basis | Functional | E$_{total}$ ground state, a.u. | E$_b$*, eV | E$_{HOMO-LUMO}$, eV | R$_{Cd-Se}$, Å | R$_{Cd-Cd}$, Å | R$_{Se-Se}$, Å |
| C$_1$ | **DFT** | **CEP-121g** | **CEP-121g** | **XA-SVWN5** | 353.1902 | **11.418** | 1.622(α) 1.216(β) | 2.714 | 2.669 | 2.342 |
| C$_1$ | DFT | LANL2DZ | LANL2DZ | B3LYP | 114.3705 | 8.223 | 2.636(α) 2.475(β) | 2.909 | 2.928 | 2.382 |
| C$_1$ | DFT | CEP-121g | CEP-121g | B3LYP | 352.3371 | 5.238 | 2.657(α) 2.555(β) | 2.848 | 2.823 | 2.350 |

*E$_b$ was evaluated for the dissociation to: Cd($^1$S) + Cd$^+$($^2$S) +2Se($^1$D)

| Ref | Electronic state | Method/Basis set | E$_b$, eV | R$_{Cd-Se}$, Å | R$_{Cd-Cd}$, Å | R$_{Se-Se}$, Å | Bond angles Se-Cd-Se, deg |
|---|---|---|---|---|---|---|---|
| **Theory** Jager M. et al J. Chem. Phys. 149 (2018) 244308 | | PBE0/cc-pVTZ-PP | | 2.76 2.76 | 2.75 | 2.21 | |

| Cd$_2$Se$_2$–(NH$_2$CH$_2$CH$_2$CH$_3$) | | | E$_t$, atomic units | R$_{Cd-Se}$, Å | R$_{Cd-Cd}$, Å | Bond angles Se-Cd-Se, deg | R$_{Cd-N}$, Å |
|---|---|---|---|---|---|---|---|
| Symmetry | Basis | Functional | | | | | |
| C$_s$ | **CEP-121g** | **XA-SVWN5** | **-386.0287** | **2.576 2.512** | **2.782** | **111.52** | **2.212** |
| C$_s$ | LANL2DZ | B3LYP | -249.8515 | 2.690 2.638 | 3.024 | 112.43 | 2.363 |
| C$_s$ | CEP-121g | B3LYP | -378.0478 | 2.650 2.587 | 2.922 | 114.21 | 2.334 |
| **Reference data** | | | | | | | |
| Crystallographic.data for Cd$_2$Se$_2$(NH$_2$CH$_2$CH$_2$CH$_3$) Yao H.-B. et al, Dalton Trans. **40** (2011) 3191 | | | | 2.591 2.581 | | | 2.162 |

Table 2. Calculation results for the Cd$_8$Se$_{13}$ clusters with different ligands

| Cluster | Symmetry Methods | Basis | | Functional | | $\Delta_{HOMO-LUMO}$, eV | $R_{Cd-Se}$, Å | $R_{Se-H}$, Å | $R_{Cd-N}$, Å | $\lambda_{max}$, nm |
|---|---|---|---|---|---|---|---|---|---|---|
| | | Geom | Abs spectra | Geom | Abs spectra | | | | | |
| Cd$_8$Se$_{13}$H$_{12}$$^{2+}$ | T$_d$ DFT | CEP-121g | LANL2DZ | XA-SVWN5 | B3LYP | 0.884 | 2.718 2.590 2.552 | 1.496 | | 404 |
| | T$_d$ DFT | LANL2DZ | | B3LYP | | 2.317 | 2.898 2.751 2.691 | 1.482 | | |
| | T$_d$ DFT | CEP-121g | | B3LYP | | 1.998 | 2.857 2.702 2.645 | 1.488 | | |
| Cd$_8$Se$_{13}$H$_{12}$ – (NH$_3$)$_4$$^{2+}$ | DFT | CEP-121g | LANL2DZ | XA-SVWN5 | B3LYP | 3.301 | 2.612 2.636 2.623 | 1.510 | 2.622 | 269 |
| Cd$_8$Se$_{13}$H$_{12}$ – (NH$_2$CH$_3$)$_4$$^{2+}$ | DFT | CEP-121g | LANL2DZ | XA-SVWN5 | B3LYP | 3.148 | 2.64 2.61 2.63 | 1.51-1.53 | 2.25 | 281 |
| Cd$_8$Se$_{13}$H$_{12}$ – (Pyr)$_4$$^{2+}$ | DFT | CEP-121g | LANL2DZ | XA-SVWN5 | B3LYP | 4.237 | 2.61-2.68 | 1.51-1.52 | 2.160 | 398 |